\PassOptionsToPackage{unicode}{hyperref}
\PassOptionsToPackage{hyphens}{url}
\PassOptionsToPackage{dvipsnames,svgnames,x11names}{xcolor}

\documentclass[
  12pt]{article}

\usepackage{amsmath,amssymb}
\usepackage{iftex}
\ifPDFTeX
  \usepackage[T1]{fontenc}
  \usepackage[utf8]{inputenc}
  \usepackage{textcomp} 
\else 
  \usepackage{unicode-math}
  \defaultfontfeatures{Scale=MatchLowercase}
  \defaultfontfeatures[\rmfamily]{Ligatures=TeX,Scale=1}
\fi
\usepackage{lmodern}
\ifPDFTeX\else  

\fi

\IfFileExists{upquote.sty}{\usepackage{upquote}}{}
\IfFileExists{microtype.sty}{
  \usepackage[]{microtype}
  \UseMicrotypeSet[protrusion]{basicmath} 
}{}
\makeatletter
\@ifundefined{KOMAClassName}{
  \IfFileExists{parskip.sty}{
    \usepackage{parskip}
  }{
    \setlength{\parindent}{0pt}
    \setlength{\parskip}{6pt plus 2pt minus 1pt}}
}{
  \KOMAoptions{parskip=half}}
\makeatother
\usepackage{xcolor}
\makeatletter
\ifx\paragraph\undefined\else
  \let\oldparagraph\paragraph
  \renewcommand{\paragraph}{
    \@ifstar
      \xxxParagraphStar
      \xxxParagraphNoStar
  }
  \newcommand{\xxxParagraphStar}[1]{\oldparagraph*{#1}\mbox{}}
  \newcommand{\xxxParagraphNoStar}[1]{\oldparagraph{#1}\mbox{}}
\fi
\ifx\subparagraph\undefined\else
  \let\oldsubparagraph\subparagraph
  \renewcommand{\subparagraph}{
    \@ifstar
      \xxxSubParagraphStar
      \xxxSubParagraphNoStar
  }
  \newcommand{\xxxSubParagraphStar}[1]{\oldsubparagraph*{#1}\mbox{}}
  \newcommand{\xxxSubParagraphNoStar}[1]{\oldsubparagraph{#1}\mbox{}}
\fi
\makeatother

\usepackage{longtable,booktabs,array}
\usepackage{calc} 
\usepackage{etoolbox}
\makeatletter
\patchcmd\longtable{\par}{\if@noskipsec\mbox{}\fi\par}{}{}
\makeatother

\IfFileExists{footnotehyper.sty}{\usepackage{footnotehyper}}{\usepackage{footnote}}
\makesavenoteenv{longtable}
\usepackage{graphicx}
\makeatletter
\def\maxwidth{\ifdim\Gin@nat@width>\linewidth\linewidth\else\Gin@nat@width\fi}
\def\maxheight{\ifdim\Gin@nat@height>\textheight\textheight\else\Gin@nat@height\fi}
\makeatother

\setkeys{Gin}{width=\maxwidth,height=\maxheight,keepaspectratio}

\makeatletter
\def\fps@figure{htbp}
\makeatother

\makeatletter
\@ifpackageloaded{caption}{}{\usepackage{caption}}
\AtBeginDocument{
\ifdefined\contentsname
  \renewcommand*\contentsname{Table of contents}
\else
  \newcommand\contentsname{Table of contents}
\fi
\ifdefined\listfigurename
  \renewcommand*\listfigurename{List of Figures}
\else
  \newcommand\listfigurename{List of Figures}
\fi
\ifdefined\listtablename
  \renewcommand*\listtablename{List of Tables}
\else
  \newcommand\listtablename{List of Tables}
\fi
\ifdefined\figurename
  \renewcommand*\figurename{Figure}
\else
  \newcommand\figurename{Figure}
\fi
\ifdefined\tablename
  \renewcommand*\tablename{Table}
\else
  \newcommand\tablename{Table}
\fi
}
\@ifpackageloaded{float}{}{\usepackage{float}}
\floatstyle{ruled}
\@ifundefined{c@chapter}{\newfloat{codelisting}{h}{lop}}{\newfloat{codelisting}{h}{lop}[chapter]}
\floatname{codelisting}{Listing}

\makeatother
\makeatletter
\@ifpackageloaded{caption}{}{\usepackage{caption}}
\@ifpackageloaded{subcaption}{}{\usepackage{subcaption}}
\makeatother

\ifLuaTeX
  \usepackage{selnolig}  
\fi
\usepackage[]{natbib}
\usepackage{bookmark}

\IfFileExists{xurl.sty}{\usepackage{xurl}}{} 
\hypersetup{
  pdftitle={Title},
  pdfauthor={Author 1; Author 2},
  pdfkeywords={3 to 6 keywords, that do not appear in the title},
  colorlinks=true,
  linkcolor={blue},
  filecolor={Maroon},
  citecolor={Blue},
  urlcolor={Blue},
  pdfcreator={LaTeX via pandoc}}

\newcommand{\anon}{1}

\begin{document}

\def\spacingset#1{\renewcommand{\baselinestretch}
{#1}\small\normalsize} \spacingset{1}

\if1\anon
{
  \title{\bf A closed form solution for Bayesian analysis of a simple linear mixed model}
  \author{Lars Erik Gangsei\\
    Faculty of Chemistry, Biotechnology and Food Science,\\ Norwegian University of Life Sciences and VI\\
    and \\
    Hilde Vinje\thanks{
    Corresponding Author, hilde.vinje@nmbu.no, Chr. M. Falsens vei 18, 1433 Ås }\hspace{.2cm} \\
    Faculty of Chemistry, Biotechnology and Food Science,\\ Norwegian University of Life Sciences}
  \maketitle
} \fi

\if0\anon
{
  \bigskip
  \bigskip
  \bigskip
  \begin{center}
    {\LARGE\bf Title}
\end{center}
  \medskip
} \fi

\bigskip
\begin{abstract}
Linear mixed-effects models are a central analytical tool for modeling hierarchical and longitudinal data, as they allow simultaneous representation of fixed and random sources of variation. In practice, inference for such models is most often based on likelihood-based approximations, which are computationally efficient, but rely on numerical integration and may be unreliable example wise  in small-sample settings. In this study, the somewhat obscure four-parameter generalized beta density is shown to be usable as a conjugate prior distribution for a simple linear mixed model. This leads to a closed-form Bayesian solution for a balanced mixed-model design, representing a methodological development beyond standard approximate or simulation-based Bayesian approaches. Although the derivation is restricted to a balanced setting, the proposed framework suggests a pathway toward analytically tractable Bayesian inference for more complex mixed-model structures. The method is evaluated through comparison with a standard frequentist solution based on likelihood estimation for linear mixed-effects models. Results indicate that the Bayesian approach performs just as well as the frequentist alternative, while yielding slightly reduced mean squared error. The study further discusses the use of empirical Bayes strategies for hyperparameter specification and outlines potential directions for extending the approach beyond the balanced case.    
\end{abstract}

\noindent
{\it Keywords:} generalized beta density, conjugate prior, random effects
\vfill

\newpage
\spacingset{1.8}

\section{Introduction}\label{sec-intro}

This study investigates the application of linear mixed-effects models to account for both fixed and random sources of variation in hierarchical or longitudinal data, under the framework outlined by \citet{bates2015lme4}. Mixed models constitute a fundamental analytical framework across a broad range of natural sciences, as they enable rigorous quantification of both fixed and random sources of variation and provide a flexible structure for modeling hierarchical, correlated, and unbalanced data, illustrated by more than 100000 citations of \citet{bates2015lme4} at scholar.google.com, medio January 2026.

When facing the problem of fitting mixed models, a common and computationally conservative approach is to use approximate methods for integrating out the random effects, estimating variance components, and performing inference. These approximate methods are widely used because they are efficient and relatively straightforward to apply. They rely on likelihood-based techniques, typically Maximum Likelihood (ML) or Restricted Maximum Likelihood (REML), to estimate model parameters. In large samples, the approximations made by these methods are generally accurate, and inference based on them tends to be reliable.

The key drawback is that these are not exact solutions. Specifically, the integration over random effects, which is essential for evaluating the likelihood, is not performed analytically, but rather approximated numerically. This can become problematic in small-sample settings, where standard errors and p-values may be biased, and approximate degrees of freedom can lead to inaccurate confidence intervals \citep{breslow1993approximate, kenward1997small}. Additionally, inference on fixed effects is often conducted by conditioning on estimated variance components, which may not fully propagate their uncertainty and can result in an underestimation of the true variability of fixed-effect estimates \citep{harville1977maximum, kenward1997small}.

In contrast, Bayesian methods provide an exact inference framework. The posterior distribution is typically not available in closed form and must be approximated numerically, typically by sampling from the true posterior distribution of all parameters, including random effects, using techniques such as Markov Chain Monte Carlo (MCMC), or other sampling methods that are outside the scope of the present study. These Bayesian methods yields a full posterior distribution that incorporates all sources of uncertainty. Bayesian methods also allow for the use of prior information, which can stabilize estimates and guide inference in the presence of sparse or noisy data. 

However, Bayesian approaches often come with their own computational challenges: they are significantly more intensive, sometimes requiring hours or longer to fit models that approximate methods can estimate in seconds \citep{spiegelhalter2002bayesian}. An exception arises when the posterior distribution has a closed form solution, as in certain conjugate models, where analytical or direct Monte Carlo sampling can be performed efficiently \citep{gelman_bayesian_2013}.

In the current study we present a distribution, the beta-gamma-normal distribution, a compounded probability distribution which is a closed form posterior distribution for the variance components and regression parameters in a simple balanced mixed model, for a particular conjugate prior distribution also presented in the current paper. We introduce a closed form solution for the posterior distribution, based on the four-parameter generalized beta density G4B ($B_{G4}$) \citep{chen1984bayesian}. 

\citet{chen1984bayesian} employ the $B_{G4}$ as the posterior distribution that arises when a three‑parameter generalized beta prior is combined with the binomial likelihood. This construction extends the classical Beta–Binomial framework by introducing additional shape parameters that allow independent control over skewness and tail behavior, thereby providing a more flexible and robust representation of posterior uncertainty than the standard two‑parameter Beta distribution. In the present study we utilize the $B_{G4}$ density as a key part of a compounded posterior distribution for a simple mixed model. To the best of our understanding and knowledge this is potentially a far more important, and novel, application beyond its somewhat specialized role as a posterior for the binomial likelihood and other reported domains of applicability.    

Furthermore, we show how empirical Bayes strategies might be applied for setting hyperparameters. Finally, we propose, without having been able to find the solution, that the principles of the current study might be extended to a more general form of mixed effect linear regression models.

\section{Methods}\label{sec-meth}

\subsection{Four-parameter generalized beta density G4B}\label{sub-sec-beta_4b_dist}
\citet{chen1984bayesian} describes the four-parameter $B_{G4}$ as a density whose probability density function is given by 

\begin{equation}\label{Eq:gen_beta_4b}
f_X(x\mid \phi_1,\phi_2,\phi_3,\delta) = \frac{1}{B\left(\phi_2,\hspace{2mm}\phi_3\right)\hspace{2mm} _{2}F_{1}\left(\phi_2,\phi_1,\phi_2+\phi_3,1-\delta\right)}\frac{x^{\phi_2-1}\left(1-x\right)^{\phi_3-1}}{\left(1-(1-\delta) x\right)^{\phi_1}}
\end{equation}

with parameters $\phi_1=\in \mathbb{R}$ and $\phi_2,\phi_3, \delta \in \mathbb{R}^+$. Further generalizations of the beta distributions is described by \citet{HAMZA201660}.

\subsection{The beta-gamma-normal ($BGN$) distribution}
Let $X_1 \in \mathbb{R},\quad 0 < X_1 < 1$, $X_2 \in \mathbb{R},\quad 0 < X_2$
and $\bf{X}_3 \in \mathbb{R}^p$ be three random variables. Let $\phi_1,\phi_2,\phi_3,\kappa_1,\kappa_2 \in \mathbb{R}^+$, with constraint $\kappa_1>\kappa_2$, $\boldsymbol{\mu}\in \mathbb{R}^p$ and $\boldsymbol{\Sigma}\in \mathbb{R}^{p\times p}$ and constraint that $\boldsymbol{\Sigma}$ is positive definite, be model parameters in the beta-gamma-normal distribution. We propose to say that $\left\{X_1,X_2,\bf{X}_3 \right\}$ is beta-gamma-normal ($BGN$) distributed, denoted $\left\{X_1,X_2,\bf{X}_3 \right\}\sim BGN\left(\phi_1,\phi_2,\phi_3,\kappa_1,\kappa_2,\boldsymbol{\mu},\boldsymbol{\Sigma}\right)$ if

\begin{equation}\label{Eq:comp_form_bgn_dist}
\begin{split}
    X_1\mid \phi_1,\phi_2,\phi_3,\kappa_1, \kappa_2 &\sim B_{G4}(\phi_1,\phi_2,\phi_3,1+\kappa_2/\kappa_1)\\
    \frac{1}{X_2}\mid X_1=x_1, \phi_1,\kappa_1,\kappa_2 & \sim Gamma(\phi_1,\hspace{2mm}\kappa_2 - \kappa_1 x_1)\\
    {\bf X}_3\mid X_1=x_1, X_2=x_2,\boldsymbol{\mu},\boldsymbol{\Sigma} & \sim N_p\left(\boldsymbol{\mu},\hspace{2mm}\frac{x_2}{(1-x_1)}\boldsymbol{\Sigma}\right)
\end{split}
\end{equation}

The latter part of this distribution, i.e. the gamma-normal distribution of $1/X_2$ and $\bf{X}_3$ is well known, i.e. \citet{gelman_bayesian_2013}. The probability density function (pdf), $f_{\left(X_1,X_2,\bf{X}_3\right)(X_1,X_2,\bf{X}_3)}$ for the $BGN$ density is given by

\begin{equation}\label{Eq:pdf_beta_gamma_normal}
\begin{split}
f_{\left(X_1,X_2,\bf{X}_3\right)}(x_1,x_2,\bf{x}_3) = &\frac{1}{B\left(\phi_3-\phi_2,\hspace{2mm}\phi_2\right)\hspace{2mm} _{2}F_{1}\left(\phi_2,\phi_1,\phi_2+\phi_3,\kappa_2/\kappa_1\right)}\frac{x_1^{\phi_2-1}\left(1-x_1\right)^{\phi_3-1}}{\left(1-(\kappa_2/\kappa_1)x_1\right)^{\phi_1}}\\
&\frac{\left(\kappa_{1}-\kappa_2 x_1\right)^{\phi_1}}{\Gamma\left(\phi_1\right)}\left(\frac{1}{x_2}\right)^{\phi_1-1}\exp\left\{-\left(\kappa_1-\kappa_2x_1\right)\frac{1}{x_2}\right\}\\
&\left(2\pi\right)^{-\frac{p}{2}} \left| \frac{x_2}{(1-x_1)} \bf{\Sigma}\right|^{-\frac{p}{2}}\exp\left\{-\frac{(1-x_1)}{2x_2} (\boldsymbol{\mu}-\bf{x}_3)^t\boldsymbol{\Sigma}^{-1}(\boldsymbol{\mu}-\bf{x}_3)\right\}
\end{split}
\end{equation}

\subsection{Closed form posterior}\label{Closed form}

A simple linear mixed model with a balanced design might be defined as

\begin{equation}\label{Eq:simple_mixed_model}
\begin{split}
y_{it} &= \bf{x}_i^t\boldsymbol{\beta} + u_i + e_{it},\hspace{2mm} i = 1,\ldots,n, \hspace{2mm} t = 1,\ldots,w\\ 
u_i&\overset{iid}{\sim} N(0,\sigma_u^2), \hspace{2mm}e_{it}\overset{iid}{\sim} N(0,\sigma^2), \hspace{2mm}cov(u_i,e_{it}) =0
\end{split}
\end{equation}

, with response variable $y_{it} \in \mathbb{R}$, predictor variable $\bf{x}_i \in \mathbb{R}^{p}$, regression parameters $\boldsymbol{\beta} \in \mathbb{R}^{p}$, random error variance $\sigma_u^2\in \mathbb{R}^+$ and random error variance $\sigma^2\in \mathbb{R}^+$. The likelihood in Eq. \ref{Eq:simple_mixed_model} might be written in its marginal form as 

\begin{equation}\label{Eq:marg_likelihood}
\bf{y} \sim N_{nw}\left(\bf{KX}\boldsymbol{\beta},\hspace{2mm} \sigma^2\boldsymbol{\Sigma}^{\otimes}\right)  
\end{equation}

, where $\bf{y} = [y_{11}\ldots y_{1w}\ldots y_{nw}]^t \in \mathbb{R}^{nw}$, $\bf{K} = \bf{1}_w\otimes \bf{I}_n$, $\bf{X} =[\bf{x}_1\ldots \bf{x}_n]^t \in \mathbb{R}^{n \times p}$ and $\boldsymbol{\Sigma}^{\otimes} = \bf{I}_n \otimes \left(\bf{I}_w + \frac{\sigma_u^2}{\sigma^2} \bf{1}_w\bf{1}_w^t\right) \in \mathbb{R}^{nw \times nw}$. Furthermore, for simplicity we define a diagonal matrix of weights, ${\bf W} = {\bf K}^t{\bf K} = w{\bf I}_n \in \mathbb{R}^{n \times n}$, a vector of mean values for the response $\overline{{\bf y}} = {\bf W}^{-1}{\bf K}^t{\bf y} \in \mathbb{R}^{n}$ and the ordinary least squares estimator for $\boldsymbol{\beta}$, $\widehat{\boldsymbol{\beta}_{ols}} = (\bf{X}^t\bf{X})^{-1}\bf{X}^t\overline{\bf{y}}$ and the two quadratic terms $Q_1 = (\bf{y}-\bf{K}\overline{\bf{y}})^t(\bf{y}-\bf{K}\overline{\bf{y}}) = \sum_{i = 1}^n\sum_{t = 1}^w(y_{it}-\overline{\bf{y}_i})^2$ and $Q_2 = (\overline{\bf{y}}-\bf{X}\widehat{\boldsymbol{\beta}_{ols}})^t\bf{W}(\overline{\bf{y}}-\bf{X}\widehat{\boldsymbol{\beta}_{ols}}) = w\sum_{i = 1}^n(\overline{\bf{y}_i}-\bf{x}_i^t\widehat{\boldsymbol{\beta}_{ols}})^2$. Finally, we define $\delta = w\sigma_u^2/(\sigma^2+w\sigma_u^2) \in \mathbb{R} \quad 0 \leq \delta < 1$. Then it might be shown, see Appendix \ref{Ap:marg_log_like} for detail, that the probability density function $f_Y(y)$ can be written as 

\begin{equation}\label{Eq:pdf_simple_mixed}
\begin{split}
f_{\bf{Y}}(\bf{y})
=& (2\pi)^{-\frac{nw}{2}}\mid\boldsymbol{\Sigma}^{\otimes}\mid^{-\frac12}\exp\left\{-\frac12\left(\bf{y}-\bf{KX}\boldsymbol{\beta}\right)^t{\boldsymbol{\Sigma}^{\otimes}}^{-1}\left(\bf{y}-\bf{KX}\boldsymbol{\beta}\right)\right\}\\
=& (2\pi)^{-\frac{nw}{2}}\left(\frac{1}{\sigma^2}\right)^{\frac{nw}{2}}\left(1-\delta\right)^{\frac{n}{2}}\\
&\exp\left\{-\frac{1}{2\sigma^2}\left[(Q_1+Q_2) - \delta Q_2 + w(1-\delta)\left(\boldsymbol{\beta}-\widehat{\boldsymbol{\beta}_{ols}}  \right)^t\bf{X}^t\bf{X}\left(\boldsymbol{\beta}-\widehat{\boldsymbol{\beta}_{ols}}  \right)\right]\right\}
\end{split}
\end{equation}

Let $\boldsymbol{\theta} = \left\{\delta, 1/\sigma^2,\boldsymbol{\beta}\right\}$ denote the unknown parameters of interest. We propose to apply the following compounded prior distribution for  $\boldsymbol{\theta}$
\begin{equation}\label{Eq:prior_dist}
\begin{split}
    \delta&\sim Beta(\nu_1\mu_1,\nu_1(1-\mu_1))\\
\frac{1}{\sigma^2} &\sim Gamma\left(\nu_2,\nu_2/\mu_2\right)\\
\boldsymbol{\beta}\mid \frac{1}{\sigma^2}, \delta&\sim N_p\left(\boldsymbol{\beta}_0,\frac{\sigma^2}{w(1-\delta)}\boldsymbol{\Upsilon}_0^{-1}\right)
\end{split}
\end{equation}

, i.e. with hyper parameters $\mu_1 \in \mathbb{R} \quad 0 < \mu_1 < 1$ and $\nu_1 \in \mathbb{R}^+$ which might be interpreted as prior mean and sample size for $\delta$, respectively. Furthermore, $\mu_2 \in \mathbb{R}^+$ and $\nu_2 \in \mathbb{R}^+$ might be interpreted as prior mean and sample size for $1/\sigma^2$. Finally, prior mean and scaled variance for $\boldsymbol{\beta}$ are given by   $\boldsymbol{\beta}_0\in \mathbb{R}^p$ and $\boldsymbol{\Upsilon}_0\in \mathbb{R}^{p\times p}$ and constraint that $\boldsymbol{\Upsilon}_0$ is positive definite. 

The pdf of the prior distribution, $f_{\boldsymbol{\Theta}}(\boldsymbol{\theta})$ is then given by

\begin{equation}\label{Eq:pdf_prior}
\begin{split}
f_{\boldsymbol{\Theta}}(\boldsymbol{\theta}) 
&= \frac{1}{B(\nu_1\mu_1,\nu_1(1-\mu_1))}\delta^{\nu_1\mu_1-1}\left(1-\delta\right)^{\nu_1(1-\mu_1)-1}\\
&\cdot \frac{(\nu_2/\mu_2)^{\nu_2}}{\Gamma(\nu_2)}\left(\frac{1}{\sigma^2}\right)^{(\nu_2)-1}exp\left\{-(\nu_2/\mu_2)\frac{1}{\sigma^2}\right\}\\
&\cdot (2\pi)^{-\frac{p}{2}}\left(w\frac{1-\delta}{\sigma^2}\right)^{\frac{p}{2}}\mid\boldsymbol{\Upsilon}_0\mid^{\frac12}exp\left\{-\frac12 w\frac{1-\delta}{\sigma^2}(\boldsymbol{\beta}-\boldsymbol{\beta}_0)^t\boldsymbol{\Upsilon}_0(\boldsymbol{\beta}-\boldsymbol{\beta}_0)\right\}
\end{split}
\end{equation}

Let the quadratic term $Q_3 = \left(\widehat{\boldsymbol{\beta}_{ols}}-\boldsymbol{\beta}_{0} \right)^tn\bf{M}_n\left(n\bf{M}_n+\boldsymbol{\Upsilon}_0\right)^{-1}\boldsymbol{\Upsilon}_0\left(\widehat{\boldsymbol{\beta}_{ols}}-\boldsymbol{\beta}_{0} \right)$, where $\bf{M}_n = \bf{X}^t\bf{X}/n$. Furthermore, let $\tilde{\boldsymbol{\beta}} = \left(n\textbf{M}_n+\boldsymbol{\Upsilon}_0\right)^{-1}\left(n\bf{M}_n\widehat{\boldsymbol{\beta}_{ols}} + \boldsymbol{\Upsilon}_0\boldsymbol{\beta}_0\right) \in \mathbb{R}^p$, $\phi_1 = nw/2+\nu_2$, $\phi_2 = \nu_1\mu_1$, $\phi_3 = n/2 +\nu_1(1-\mu_1)$, $\kappa_1 = \frac12(Q_1+Q_2+2\nu_2/\mu_2+wQ_3)$ and $\kappa_2 = \frac12(Q_2+wQ_3)$, all $\in \mathbb{R}^+$ . 

When the prior distribution in Eq. \ref{Eq:prior_dist} is applied to the model defined in Eq. \ref{Eq:simple_mixed_model} the posterior distribution for $\boldsymbol{\theta}$ is given by a beta-gamma-normal distribution, i.e.

\begin{equation}
\left\{\delta,1/\sigma^2,\boldsymbol{\beta}\mid \bf{y} \right\}\sim BGN\left(\phi_1,\phi_2,\phi_3,\kappa_1,\kappa_2,\tilde{\boldsymbol{\beta}},\frac{1}{w}\left[n\bf{M}_n+\boldsymbol{\Upsilon}_0\right]^{-1}\right)
\end{equation}

, or in the compounded form the posterior might be written as:

\begin{equation}\label{Eq:comp_form_posterior}
\begin{split}
    \delta &\sim B_{G4}\left(\phi_1, \phi_2,\phi_3,1+\frac{\kappa_2}{\kappa_1}\right)\\
    \frac{1}{\sigma^2}\mid \delta & \sim Gamma\left(\phi_1,\hspace{2mm}\kappa_1-\kappa_2\delta \right)\\
    {\boldsymbol \beta}\mid \delta, \frac{1}{\sigma^2}& \sim N_p\left(\tilde{\boldsymbol{\beta}},\hspace{2mm}\frac{\sigma^2}{w(1-\delta)}\left[n\bf{M}_n+\boldsymbol{\Upsilon}_0\right]^{-1}\right)
\end{split}
\end{equation}

Proof: 
Since 
\begin{equation}\label{Eq:proof_posterior}
\begin{split}
    f_\Theta(\theta\mid \Phi)\cdot f_{\textbf{Y}}(\textbf{y}) = \mathit{C} &\cdot \frac{1}{B\left(\phi_2,\phi_3\right) {_2F_1}\left(\phi_2,\phi_1,\phi_2+\phi_3,\frac{\kappa_2}{\kappa_1}\right)}\\
    &\cdot \frac{\delta^{\phi_2-1}\cdot\left(1-\delta\right)^{\phi_3-1}}{\left[\left(1 -(\kappa_2/\kappa_1)\delta\right)\right]^{\phi_1}}\\
&\cdot \frac{\left[\left(\kappa_1-\kappa_2\delta\right)\right]^{\phi_1}}{\Gamma(\phi_1)}\left(\frac{1}{\sigma^2}\right)^{\phi_1-1}exp\left\{\left(\kappa_1-\kappa_2\delta\right)\frac{1}{\sigma^2}\right\}\\
&\cdot (2\pi)^{-\frac{p}{2}}\left(\frac{1-\delta}{\sigma^2}\right)^{\frac{p}{2}}\mid w\left(n\bf{M}_n+\boldsymbol{\Upsilon}_0\right)\mid^{\frac12}\\
&\hspace{3mm}exp\left\{-\frac12\left[\left(\boldsymbol{\beta}-\widetilde{\boldsymbol{\beta}}\right)^t\frac{1-\delta}{\sigma^2}w\left(n\bf{M}_n+\boldsymbol{\Upsilon}_0\right)\left(\boldsymbol{\beta}-\widetilde{\boldsymbol{\beta}}\right)\right]\right\}
    \end{split}
\end{equation}

, where $\mathit{C}$ is a constant independent of $\boldsymbol{\theta}$, i.e.

\begin{equation}\label{Eq:model_evidence}
\mathit{C} = (2\pi)^{-\frac{nw}{2}}\frac{B\left(\phi_2,\phi_3\right) {_2F_1}\left(\phi_2,\phi_1,\phi_2+\phi_3,\frac{\kappa_2}{\kappa_1}\right)}{B(\nu_1\mu_1,\nu_1(1-\mu_1))}\cdot\frac{(\nu_2/\mu_2)^{\nu_2}}
{\kappa_1^{\phi_1}} \cdot\frac{\mid\boldsymbol{\Upsilon}_0\mid^{\frac12}}{\mid n\boldsymbol{M}_n+\boldsymbol{\Upsilon}_0\mid^{\frac12}}\cdot\frac{\Gamma(\phi_1)}{\Gamma(\nu_2)}
\end{equation}

, we see that the kernel in Eq. \ref{Eq:proof_posterior} represents probability density function from a BGN distribution, see Eq. \ref{Eq:pdf_beta_gamma_normal}.

\subsection{Empirical Bayes and model evidence}
Since Eq. \ref{Eq:proof_posterior} might be written as $f_\Theta(\theta\mid \boldsymbol{\Phi})\cdot f_{\textbf{Y}}(\textbf{y}) = \mathit{C} \cdot f_{\Theta\mid \textbf{Y}}(\theta\mid \textbf{Y}, \boldsymbol{\Phi})$, where $f_{\Theta\mid \textbf{Y}}(\theta\mid \textbf{Y}, \boldsymbol{\Phi})$ is the pdf of the posterior distribution,  we know from Bayes theorem that $ \mathit{C} = \int_\theta f_\Theta(\theta\mid \boldsymbol{\Phi})\cdot f_{\textbf{Y}}(\textbf{y})\partial \theta$, also known as "model evidence". An approximate interpretation is, "probability of data given model", consequently the model evidence might be used for Bayesian model selection \citealp{gelman_bayesian_2013}. As the model evidence is a function of prior hyperparameters, i.e. $\boldsymbol{\Phi}$, one method within the "Empirical Bayes" framework is to maximize the model evidence with respect to hyper-parameters, i.e. $\boldsymbol{\Phi}_{EB} = \underset{\Phi}{argmax} \hspace{1mm}C(\boldsymbol{\Phi},\textbf{Y})$. 

For practical purposes some restrictions regarding prior hyperparameters, $\boldsymbol{\Phi}$. One natural choice is to use Zellners G prior \citep{zellner1986gprior}, i.e. $\boldsymbol{\Upsilon}_0 = (\nu_3 \textbf{M}_n)^{-1}$. Furthermore, we regard $\mu_1$, $\mu_2$ and $\boldsymbol{\beta}_0$ as known an fixed hyperparameters. In the accompanying R package \texttt{bmmix} \citep{vinje_bmmix_2025} the default values are set till $\mu_1 = \frac12$, $\mu_2 = 1$ and $\boldsymbol{\beta}_0 = \textbf{0}_p$, but any valid values might be chosen. Finally, the three free prior hyper parameters, $\nu_1$, $\nu_2$ and $\nu_3$, might be set by $\left\{\nu_1,\nu_2,\nu_3\right\} = \underset{\nu_1,\nu_2,\nu_3}{argmax}\hspace{2mm} \ell(C\mid \mu_1,\mu_2,\boldsymbol{\beta}_0)$, where 

\begin{equation}\label{Eq:log_evidence}
\begin{split}
    \ell(C\mid \mu_1,\mu_2,\boldsymbol{\beta}_0) = &-\frac{nw}{2}log(2\pi) + log\left\{{_2F_1}\left(\nu_1\mu_1,\frac{nw}{2}+\nu_2,\frac{n}{2}+\nu_1,\frac{\kappa_2}{\kappa_1}\right)\right\}\\
    &+\nu_2\left[log(\nu_2)-\log(\mu_2)\right] - (\frac{nw}{2}+\nu_2)log(\kappa_1) +\frac{p}{2} \left[log(\nu_3) -log(n +\nu_3)\right]\\
    &+log\left[\Gamma(\frac{nw}{2}+\nu_2)\right]
    -log\left[\Gamma(\nu_2)\right]
    -log\left[\Gamma(\frac{n}{2}+\nu_1)\right]\\
    &+log\left[\Gamma(\frac{n}{2}+\nu_1(1-\mu_1))\right]
     +log\left[\Gamma(\nu_1)\right] 
     -log\left[\Gamma(\nu_1(1-\mu_1))\right] 
\end{split}
\end{equation}

It should be noticed that $\kappa_1$ and $\kappa_2$ are functions of $\nu_2$, $\mu_2$, $\nu_3$ and $\boldsymbol{\beta}_0$, and that due to Zellner's g prior $Q_3$ simplifies till $Q_3 = \frac{\nu_3}{n+\nu_3}(\widehat{\boldsymbol{\beta}_{ols}}-\boldsymbol{\beta}_0)^tn\textbf{M}_n(\widehat{\boldsymbol{\beta}_{ols}}-\boldsymbol{\beta}_0)$. In the R package it is included a non-linerar optimizer for $\nu_1$, $\nu_2$ and $\nu_3$.

\subsection{Data simulation}
To compare the new Bayesian inference method with the conventional Frequentist inference approach, we simulated 1000 independent datasets using a repeated-measures structure. For each simulation replicate, we generated a unique random seed to ensure reproducibility and independence between runs. We set n = 20, w = 4 and p = 3 and parameters $\sigma^2 = 4$, $\sigma_u^2 = 0.5$ and $\boldsymbol{\beta} = [0.2\hspace{2mm} 2 \hspace{2mm} -0.5]^t$ fixed across simulations.  

For every replicate, we constructed a new design matrix, $\textbf{X}\in \mathbb{R}^{20\times 3}$, with the first column (intercept) as $\textbf{1}_{20}$ and the remaining 40 elements as independent standard normally distributed elements representing two covariates. Subject-specific random effects were generated and applied across repeated measurements, and independent normally distributed error terms were added to represent measurement noise. The final outcome values were obtained by combining the fixed effects, random effects, and measurement error. All simulated values, covariates, and subject identifiers were stored in long format for model fitting and comparison. 

An implementation of this simulation setup is included as an example in the \texttt{bmlmer} (\texttt{bmmix}) function in R \citep{vinje_bmmix_2025}.

We compared the performance of the proposed closed-form Bayesian posterior with that of the standard frequentist mixed-effects modeling approach implemented in \texttt{lmer} (\texttt{lme4}) function in R, see \citet{bates2015lme4} for details. We used the empirical Bayes setup outlined above to set prior sample sizes $\nu_2$ and $\nu_3$, whereas $\nu_1$, i.e. prior sample size for $\delta$ was restricted to to the interval $[2\hspace{2mm}2.001]$ which is an equivalent of the uniform prior for $\delta$ for the parametrization of the beta distribution applied in this study.  

We calculated the respective 95\% credibility intervals and approximate confidence intervals. We apply the abbreviation CI for both kind of intervals for simplicity. We report the "overlap", i.e. the proportion of simulations that the true parameter value is covered by the associated CI, and the average "width" of CI's. Ideally the overlap should be approximately 0.95 for 95\% CI's and the "witdh" as small/ narrow as possible. 

Furthermore, we report mean square error "MSE" as $\frac{1}{1000}\sum_{i=1}^{1000}(\widehat{\theta}-\theta)^2$, and "bias" as $\frac{1}{1000}\sum_{i=1}^{1000}(\widehat{\theta}-\theta)$, where $\theta$ is used as an alias for all five/ six (as we also report for $\delta$ in the Bayesian framework) parameters in question. Posterior means and maximum likelihood estimates were used as parameter estimates , $\widehat{\theta}$'s, in the Bayesian and frequentist framework respectively.

\subsection{Future perspectives}
Considerable effort has been allocated to finding a closed form solution for the posterior distribution for the more general form of the mixed model, i.e.  

\begin{equation}\label{Eq:mixed_model}
\begin{split}
y_{it} &= \bf{x}_i^t\boldsymbol{\beta} + \bf{z}_i^t\bf{u}_i + e_{it},\hspace{2mm} i = 1,\ldots,n, \hspace{2mm} t = 1,\ldots,w_i\\ 
\bf{u}_i&\overset{iid}{\sim} N_q(0,\sigma^2\boldsymbol{\Lambda}), \hspace{2mm}e_{it}\overset{iid}{\sim} N(0,\sigma^2)
\end{split}
\end{equation}

The model presented in Eq. (\ref{Eq:mixed_model}) might be unbalanced (i.e. different number of observations $w_i$ for each level $i$). Furthermore, for each level the random effect is given by vector product $\bf{z}_i^t\bf{u}_i$ where $\bf{z}_i \in \mathbb{R}^q$ is a design vector and $\bf{u}_i \in \mathbb{R}^q$ is the random regression parameters assumed to be normal with zero mean and constant covariance matrix, $\sigma^2\boldsymbol{\Lambda}\in \mathbb{R}^{q\times q}$.

Let $N = \sum_{i=1}^nw_i$ be the total number of observations, $\textbf{y}_i = [y_{i1}\ldots y_{iw_i}]^t\in \mathbb{R}^{w_i}$ , $\textbf{y} = [\textbf{y}_{1}^t\ldots \textbf{y}_{n}^t]^t\in \mathbb{R}^{N}$, the diagonal block matrix $\textbf{K} \in \mathbb{R}^{N \times n}$ with zeros and ones so that the column wise sums are equal to $w_i$ and $\textbf{W} = \textbf{K}^t\textbf{K} \in \mathbb{R}^{n \times n}$ is a diagonal matrix with main diagonal $[w_1 \ldots w_n]^t$, and finally let $\textbf{Z} \in \mathbb{R}^{nq \times n}$ be a block diagonal matrix where the vectors $\textbf{z}_1\ldots \textbf{z}_n$ constitutes the block elements. 

With this setup the marginal distribution of $\textbf{y}$ given in Eq. \ref{Eq:marg_likelihood}, is still valid, but we have  

\begin{equation}\label{Eq:sigma_kron}
\begin{split}
\boldsymbol{\Sigma}^{\otimes} &= \textbf{I}_N + \textbf{KZ}^t\left(\textbf{I}_n \otimes \boldsymbol{\Lambda}\right)\textbf{ZK}^t
\\
{\boldsymbol{\Sigma}^{\otimes}}^{-1} &= \textbf{I}_N - \textbf{KZ}^t\left(\textbf{I}_n \otimes \boldsymbol{\Lambda}^{-1}+\textbf{ZWZ}^t\right)^{-1}\textbf{ZK}^t\\
\mid{\boldsymbol{\Sigma}^{\otimes}}\mid &= \mid \boldsymbol{\Lambda}\mid^n\mid\textbf{I}_n \otimes \boldsymbol{\Lambda}^{-1}+\textbf{ZWZ}^t\mid
\end{split}
\end{equation}

The inverse form in Eq. \ref{Eq:sigma_kron} is easily derived by the Woodbury matrix identity and the determinant by the matrix determinant lemma. It might be noted that $\textbf{KZ}^t\left(\textbf{I}_n \otimes \boldsymbol{\Lambda}\right)\textbf{ZK}^t \in \mathbb{R}^{N\times N}$ is block diagonal, where the $i$th block is given by $(\textbf{z}_i^t\boldsymbol{\Lambda}\textbf{z}_i)\textbf{1}_{w_i}\textbf{1}_{w_i}^t$. 

Furthermore, let the diagonal matrix $\textbf{G} \in \mathbb{R}^{n \times n}$, the quadratic term $Q_2 \in \mathbb{R}^+$ and $(1 - \delta) \in \mathbb{R}$ so that $0<(1 -\delta)<1$ be given by 

\begin{equation}\label{Eq:definitions}
\begin{split}
    \textbf{G} &= \textbf{WZ}^t\left(\textbf{I}_n \otimes \boldsymbol{\Lambda}^{-1}+\textbf{ZWZ}^t\right)^{-1}\textbf{ZW} = \textbf{W} - \textbf{K}^t{\boldsymbol{\Sigma}^{\otimes}}^{-1}\textbf{K} \in \mathbb{R}^{n\times n}\\
    &\Rightarrow \textbf{K}^t{\boldsymbol{\Sigma}^{\otimes}}^{-1}\textbf{K} = \textbf{W} -\textbf{G}\\
    \overline{\textbf{y}} &= \textbf{W}^{-1}\textbf{K}^t\textbf{y}\in \mathbb{R}^n\\
    \widehat{\boldsymbol{\beta}_{\boldsymbol{\Lambda}}} &= \left(\textbf{X}^t\textbf{K}^t{\boldsymbol{\Sigma}^{\otimes}}^{-1}\textbf{K}\textbf{X}\right)^{-1}\textbf{X}^t\textbf{K}^t{\boldsymbol{\Sigma}^{\otimes}}^{-1}\textbf{K}\overline{\textbf{y}}\in \mathbb{R}^p\\
    \widehat{\boldsymbol{\beta}_{w}} &= \left(\textbf{X}^t\textbf{W}\textbf{X}\right)^{-1}\textbf{X}^t\textbf{W}\overline{\textbf{y}}\in \mathbb{R}^p\\
    Q_1 &= \sum_{i=1}^n\sum_{t= 1}^{w_i}(y_{it} -\overline{y_i})^2 \in \mathbb{R}^+\\
    Q_2 &= \left(\overline{\textbf{y}}-\textbf{X}\widehat{\boldsymbol{\beta}_{w}}\right)^t\textbf{W}\left(\overline{\textbf{y}}-\textbf{X}\widehat{\boldsymbol{\beta}_{w}}\right) \in \mathbb{R}^{+}\\
    (1-\delta) &= \left[1-\frac{1}{Q_2}\left(\overline{\textbf{y}}-\textbf{X}\widehat{\boldsymbol{\beta}_{w}}\right)^t\textbf{G}\left(\overline{\textbf{y}}-\textbf{X}\widehat{\boldsymbol{\beta}_{w}}\right)\right]^{\frac1n} \in \mathbb{R}, 0\leq \delta\leq 1
    \end{split}
\end{equation} 

Then, see Appendix \ref{Ap:marg_log_like} and Appendix \ref{Ap:more_calc} for details, the marginal pdf, $f_\textbf{Y}(\textbf{y};\sigma^2, \boldsymbol{\Lambda},\boldsymbol {\beta})$ might be written as 

\begin{equation}\label{Eq:marg_pdf}
\begin{split}
f_\textbf{Y}(\textbf{y};\sigma^2, \boldsymbol{\Lambda},\boldsymbol {\beta}) = (2\pi)^{-\frac{N}{2}}\mid\sigma^2\boldsymbol{\Sigma}^{\otimes}\mid^{-\frac12}\exp&\left\{-\frac{1}{2\sigma^2}\left(\textbf{y}-\textbf{KX}\boldsymbol{\beta}\right)^t{\boldsymbol{\Sigma}^{\otimes}}^{-1}\left(\textbf{y}-\textbf{KX}\boldsymbol{\beta}\right)\right\}\\
= (2\pi)^{-\frac{N}{2}}\mid\sigma^2\boldsymbol{\Sigma}^{\otimes}\mid^{-\frac12}\exp&\left\{-\frac{1}{2\sigma^2}\left[Q_1\right.\right.\\
&+\left(\overline{\textbf{y}}-\textbf{X}\widehat{\boldsymbol{\beta}_{w}}\right)^t[\textbf{W}-\textbf{G}]\left(\overline{\textbf{y}}-\textbf{X}\widehat{\boldsymbol{\beta}_{w}}\right)\\
&- \left(\widehat{\boldsymbol{\beta}_{w}}-\widehat{\boldsymbol{\beta}_{\boldsymbol{\Lambda}}}\right)^t\textbf{X}^t[\textbf{W}-\textbf{G}]\textbf{X}\left(\widehat{\boldsymbol{\beta}_{w}}-\widehat{\boldsymbol{\beta}_{\boldsymbol{\Lambda}}}\right)\\
&+\left.\left.
\left(\boldsymbol{\beta}-\widehat{\boldsymbol{\beta}_{\boldsymbol{\Lambda}}}\right)^t\textbf{X}^t\textbf{K}^t{\boldsymbol{\Sigma}^{\otimes}}^{-1}\textbf{K}\textbf{X}\left(\boldsymbol{\beta}-\widehat{\boldsymbol{\beta}_{\boldsymbol{\Lambda}}}\right)\right]\right\}\\
= (2\pi)^{-\frac{N}{2}}\mid\sigma^2\boldsymbol{\Sigma}^{\otimes}\mid^{-\frac12}\exp&\left\{-\frac{1}{2\sigma^2}\left[Q_1 + Q_2(1-\delta)^n\right.\right.\\
&-\text{trace}\left(\left[\widehat{\boldsymbol{\beta}_{w}}-\widehat{\boldsymbol{\beta}_{\boldsymbol{\Lambda}}}\right]\left[\widehat{\boldsymbol{\beta}{w}}-\widehat{\boldsymbol{\beta}_{\boldsymbol{\Lambda}}}\right]^t\textbf{X}^t[\textbf{W}-\textbf{G}]\textbf{X}\right)\\
&+\left.\left.
\left(\boldsymbol{\beta}-\widehat{\boldsymbol{\beta}_{\boldsymbol{\Lambda}}}\right)^t\textbf{X}^t\textbf{K}^t{\boldsymbol{\Sigma}^{\otimes}}^{-1}\textbf{K}\textbf{X}\left(\boldsymbol{\beta}-\widehat{\boldsymbol{\beta}_{\boldsymbol{\Lambda}}}\right)\right]\right\}
\end{split}
\end{equation}

In Eq. \ref{Eq:marg_pdf} $Q_1$ and $Q_2$ are just a function of data and the term $\left(\boldsymbol{\beta}-\widehat{\boldsymbol{\beta}_{\boldsymbol{\Lambda}}}\right)^t\textbf{X}^t\textbf{K}^t{\boldsymbol{\Sigma}^{\otimes}}^{-1}\textbf{K}\textbf{X}\left(\boldsymbol{\beta}-\widehat{\boldsymbol{\beta}_{\boldsymbol{\Lambda}}}\right)$ has its desired form. 
Also note that $Q_2$ is just a generalization of the formula used for the simple balanced design. 

Furthermore, for the simple balanced design with $\textbf{W} = w\textbf{I}_n$, $\textbf{Z} = \textbf{I}_n$ and $\boldsymbol{\Lambda} = \frac{\sigma_u^2}{\sigma^2}$ we get $\delta = \frac{w\sigma_u^2}{\sigma^2+w\sigma_u^2}$ applying the generalization for $(1-\delta)$ in Eq. \ref{Eq:definitions}, which is in line with the definition of $\delta$ used earlier in this paper. 

We notice the following:  

1) In the simple balanced case it is "easy" to calculate $\widehat{\boldsymbol{\beta}}$ since $\widehat{\boldsymbol{\beta}_{\boldsymbol{\Lambda}}} = \widehat{\boldsymbol{\beta}_{ols}} = \widehat{\boldsymbol{\beta}_{w}}$, where  $\widehat{\boldsymbol{\beta}_{w}} = \left(\textbf{X}^t\textbf{W}\textbf{X}\right)^{-1}\textbf{X}^t\textbf{W}\overline{\textbf{y}}\in \mathbb{R}^p$

2) It seems clear that $\underset{\frac{\mid\boldsymbol{\Lambda}\mid}{\sigma^2}\to 0}{\text{lim } \widehat{\boldsymbol{\beta}_{\boldsymbol{\Lambda}}}} = \widehat{\boldsymbol{\beta}_{w}}$. 

3) If $\textbf{C} = \textbf{X}^t\textbf{WZ}^t\left[\left(\textbf{I}_n\otimes \boldsymbol{\Lambda}^{-1}\right)+\textbf{ZW}\left(\textbf{I}-\textbf{X}(\textbf{X}^t\textbf{W}\textbf{X}^t)^{-1}\textbf{X}^t\textbf{W}\right)\textbf{Z}^t\right]^{-1}\textbf{ZWX}$, in $\mathbb{R}^{p\times p}$, then by the Woodbury matrix identity $\left(\textbf{X}^t\textbf{K}^t{\boldsymbol{\Sigma}^{\otimes}}^{-1}\textbf{K}\textbf{X}\right)^{-1} = \left(\textbf{X}^t\textbf{WX}\right)^{-1} + \left(\textbf{X}^t\textbf{WX}\right)^{-1}\textbf{C}\left(\textbf{X}^t\textbf{WX}\right)^{-1}$.

4) We suspect that there might be a solution with a compounded distribution where $\delta$ (which might well have another form than suggested above) is $B_{G4}$, $1/\sigma^2\mid \delta$ is gamma distributed, $\boldsymbol{\Lambda}^{-1}\mid \delta, \sigma^2$ or some function of $\boldsymbol{\Lambda}^{-1}$ is Wishart distributed and finally $\boldsymbol{\beta}\mid \boldsymbol{\Lambda},\delta, \sigma^2$ is normal. It might also be possible that $\widehat{\boldsymbol{\beta}_{\boldsymbol{\Lambda}}}$ might be possible to calculate if $\delta$ and $\sigma^2$ is known.

\section{Results}\label{sec-res}

The contribution of this article lies primarily in methodological development. The proposed method and distributions are implemented and made publicly available through the newly developed R package \texttt{bmmix} \citep{vinje_bmmix_2025}. The \texttt{bmmix} package relies in the \texttt{gbeta} package \citep{laurent2020gbeta}, see also \citet{laurent2012some},  which provides a comprehensive set of tools for working with the $B_{G4}$ distribution, including functions for density evaluation, distribution and quantile functions, and random number generation. The \texttt{bmmix} package, implements a closed-form Bayesian posterior for simple, balanced mixed-effects linear regression models based on the Extended Beta Distribution of the third kind. A function for utilizing empirical Bayes is also included in the package. 

The comparative results between the Bayesian approach with the empirical Bayes priors, and the frequentist are presented in Table 1. It summarises the results from 1000 independent simulations ($n = 20$, $w = 4$, $p = 3$). Overall, the Bayesian and frequentist methods perform similarly across all parameters, with comparable coverage probabilities and small biases. For the fixed-effect parameters ($\beta_0$, $\beta_1$, $\beta_2$), both approaches show coverage close to the nominal 95\% level and minimal bias. As expected a priori the Bayesian regression parameters are shrinked towards 0. Furthermore, for $\sigma_u^2$ the Bayesian CI's are wider than the frequentist counterpart, and probably "too wide" with a 99\% overlap. On the other hand, MSE for the Bayesian alternative (0.14) is considerably lower than the frequentist alternative (0.24). These aspects are not evaluated in detail, however we noticed that the maximum likelihood frequentist estimate for $\sigma_u^2$ was equal to 0 in a substantial proportion of the simulations. In terms of mean squared error (MSE), the Bayesian estimates tend to have equal or lower values than the frequentist estimates for most parameters, indicating a modest improvement in estimator precision. 

Taken together, the results suggest that the Bayesian method performs at least as well as the frequentist approach in this setting, with slight advantages in MSE for several parameters.

\begin{table}[ht]
\caption{Results from simulation study, based on results from 1000 independent simulations, with n = 20, w = 4, p = 3. Left column, parameter names, second left column, true parameter values. Then for the Bayesian and frequentist inferences "Overlap", i.e. the proportion of simulations for when true parameter value was covered by a 95\% CI, "Width"; average width of CI's, "MSE"; mean square error, i.e. $\overline{(\widehat{\theta}-\theta)^2}$ and "Bias", i.e. $\overline{(\widehat{\theta}-\theta)}$.}
\centering
\begin{tabular}{|c|r|rrrr|rrrr|}
  \hline  
  \multicolumn{2}{|c|}{}&\multicolumn{4}{|c}{Bayesian inference}&\multicolumn{4}{|c|}{Frequentist inference}\\
  \hline
 Parameter & True & Overlap & Width & MSE & Bias & Overlap & Width & MSE & Bias \\ 
  \hline
$\delta$ & 0.33 & 0.98 & 0.60 & 0.01 & 0.01 &  &  &  &  \\ 
  $\sigma^2$ & 4.00 & 0.94 & 2.66 & 0.46 & -0.06 & 0.95 & 2.75 & 0.49 & -0.07 \\ 
  $\sigma_u^2$ & 0.50 & 0.99 & 1.81 & 0.14 & 0.16 & 0.95 & 1.59 & 0.24 & 0.04 \\  
  $\beta_0$ & 0.20 & 0.96 & 1.15 & 0.07 & -0.02 & 0.94 & 1.10 & 0.08 & -0.01 \\  
  $\beta_1$ & 2.00 &  0.91 & 1.20 & 0.10 & -0.13 & 0.93 & 1.15 & 0.09 & -0.01 \\ 
  $\beta_2$ & -0.50 & 0.96 & 1.19 & 0.08 & 0.03 & 0.93 & 1.14 & 0.09 & 0.00 \\ 
   \hline
\end{tabular}
\end{table}

\section{Conclusion}\label{sec-conc}
An exact solution for fitting mixed-effects models is advantageous in light of the limitations associated with approximate likelihood-based methods, particularly in settings with small sample sizes or complex variance structures, as discussed in the introduction. In this article, we have taken a step toward addressing these challenges by deriving a closed-form Bayesian posterior for a class of balanced mixed-effects linear models based on the Extended Beta Distribution of the third kind.

The proposed methodology accounts for uncertainty through an exact closed-form Bayesian posterior, without relying on computationally intensive simulation-based Bayesian methods. For a simple, balanced mixed-effects linear regression model, this provides a fully analytical alternative to approximate likelihood-based approaches. To facilitate practical use and reproducibility, the methods have been implemented and made publicly available through a dedicated R package, enabling straightforward application and comparison with established methods in applied settings.

The results presented here demonstrate that an exact Bayesian solution can be derived for a simple balanced mixed-effect linear regression mode, and illustrate the potential of this approach. While further work is required to extend the methodology to more general model formulations, we believe that the framework introduced in this study may be of interest for continued methodological development.

\section{Disclosure statement}\label{disclosure-statement}

The authors have no conflicts of interest to declare.

\section{Data Availability Statement}\label{data-availability-statement}

All data used in this study are simulated and can be fully reproduced using the example code provided in the help documentation for the \texttt{bmlmer} function in the \texttt{bmmix} R package \citep{vinje_bmmix_2025}.

\bibliography{GangseiVinjeBibliography.bib}

\clearpage
\appendix
\section{Appendix 1: Quadratic form of the log likelihood}\label{Ap:marg_log_like}
The marginal log likelihood for the general linear mixed model defined Eq. \ref{Eq:marg_likelihood} is given by Eq. \ref{Eq:marg_pdf}. Calculations below show how the quadratic form in Eq. \ref{Eq:quad_form_general} is derived. 

\begin{equation}\label{Eq:quad_form_general}
\begin{split}
\left(\textbf{y}-\textbf{KX}\boldsymbol{\beta}\right)&^t{\boldsymbol{\Sigma}^{\otimes}}^{-1}\left(\textbf{y}-\textbf{KX}\boldsymbol{\beta}\right)\\
= &\left(\textbf{y}-\textbf{K}\overline{\textbf{y}} + \textbf{K}\overline{\textbf{y}}- \textbf{KX}\boldsymbol{\beta}\right)^t{\boldsymbol{\Sigma}^{\otimes}}^{-1}\left(\textbf{y}-\textbf{K}\overline{\textbf{y}} + \textbf{K}\overline{\textbf{y}}- \textbf{KX}\boldsymbol{\beta}\right)\\
= &\underset{=Q_1\text{ since } \left(\textbf{y}-\textbf{K}\overline{\textbf{y}} \right)^t\textbf{K} = \textbf{y}^t\left(\textbf{I}_N - \textbf{K}(\textbf{K}^t\textbf{K})^{-1}\textbf{K}\right)\textbf{K} = \textbf{y}^t\textbf{0}_N\textbf{0}_n^t}{\underbrace{\left(\textbf{y}-\textbf{K}\overline{\textbf{y}} \right)^t{\boldsymbol{\Sigma}^{\otimes}}^{-1}\left(\textbf{y}-\textbf{K}\overline{\textbf{y}} \right)}}\\
&+ 2 \underset{= 0 \text{ since the term } \left(\textbf{y}-\textbf{K}\overline{\textbf{y}} \right)^t\textbf{K} \text{ becomes a part of both elements in }{\boldsymbol{\Sigma}^{\otimes}}^{-1}}{\underbrace{\left(\textbf{y}-\textbf{K}\overline{\textbf{y}} \right)^t{\boldsymbol{\Sigma}^{\otimes}}^{-1}\textbf{K}\left(\overline{\textbf{y}}- \textbf{X}\boldsymbol{\beta}\right)}}\\
&+ \left(\overline{\textbf{y}}- \textbf{X}\underset{=\widehat{\boldsymbol{\beta}_\Lambda}}{\underbrace{\left(\textbf{X}^t\textbf{K}^t{\boldsymbol{\Sigma}^{\otimes}}^{-1}\textbf{K}\textbf{X}\right)^{-1}\textbf{X}^t\textbf{K}^t{\boldsymbol{\Sigma}^{\otimes}}^{-1}\textbf{K}\overline{\textbf{y}}}} +\textbf{X}\widehat{\boldsymbol{\beta}_\Lambda}- \textbf{X}\boldsymbol{\beta} \right)^t\textbf{K}^t{\boldsymbol{\Sigma}^{\otimes}}^{-1}\textbf{K}\left(\overline{\textbf{y}}- \textbf{X}\widehat{\boldsymbol{\beta}_\Lambda} +\ldots \right)\\
=  &Q_1+ \overline{\textbf{y}}^t\left(\textbf{I}_n-\underset{\textbf{H}_\Lambda^t}{\underbrace{\textbf{K}^t{\boldsymbol{\Sigma}^{\otimes}}^{-1}\textbf{K}\textbf{X}\left(\textbf{X}^t\textbf{K}^t{\boldsymbol{\Sigma}^{\otimes}}^{-1}\textbf{K}\textbf{X}\right)^{-1}\textbf{X}^t}}\right)\textbf{K}^t{\boldsymbol{\Sigma}^{\otimes}}^{-1}\textbf{K}\left(\textbf{I}_n-\textbf{H}_\Lambda\right)\overline{\textbf{y}}\\
+ &2\underset{= 0 \text{ since } \left(\textbf{I}_n-\textbf{H}_\Lambda^t\right)\textbf{K}^t{\boldsymbol{\Sigma}^{\otimes}}^{-1}\textbf{K}\textbf{X} = \textbf{0}_n\textbf{0}_n^t}{\underbrace{\overline{\textbf{y}}^t\left(\textbf{I}_n-\textbf{H}_\Lambda^t\right)\textbf{K}^t{\boldsymbol{\Sigma}^{\otimes}}^{-1}\textbf{K}\textbf{X}\left(\widehat{\boldsymbol{\beta}_\Lambda} - \boldsymbol{\beta}\right)}}\\
+& \left(\boldsymbol{\beta}-\widehat{\boldsymbol{\beta}_{\boldsymbol{\Lambda}}}\right)^t\textbf{X}^t\textbf{K}^t{\boldsymbol{\Sigma}^{\otimes}}^{-1}\textbf{K}\textbf{X}\left(\boldsymbol{\beta}-\widehat{\boldsymbol{\beta}_{\boldsymbol{\Lambda}}}\right)\\
= &Q_1+
\left(\overline{\textbf{y}}-\textbf{X}\widehat{\boldsymbol{\beta}_{\boldsymbol{\Lambda}}}\right)^t\textbf{K}^t{\boldsymbol{\Sigma}^{\otimes}}^{-1}\textbf{K}\left(\overline{\textbf{y}}-\textbf{X}\widehat{\boldsymbol{\beta}_{\boldsymbol{\Lambda}}}\right)
+\left(\boldsymbol{\beta}-\widehat{\boldsymbol{\beta}_{\boldsymbol{\Lambda}}}\right)^t\textbf{X}^t\textbf{K}^t{\boldsymbol{\Sigma}^{\otimes}}^{-1}\textbf{K}\textbf{X}\left(\boldsymbol{\beta}-\widehat{\boldsymbol{\beta}_{\boldsymbol{\Lambda}}}\right)
\end{split}
\end{equation}

For the simple, balanced design, defined in Eq. \ref{Eq:simple_mixed_model} might be seen as a generalization of Eq. \ref{Eq:mixed_model}, where 
we have that $\textbf{K} = \textbf{I}_n\otimes \textbf{1}_w$, $\textbf{Z} = \textbf{I}_n$ and $\boldsymbol{\Lambda} = \frac{\sigma_u^2}{\sigma^2}$ and consequently ${\boldsymbol{\Sigma}^{\otimes}}^{-1} = \textbf{I}_N - \frac{\sigma_u^2}{\sigma^2+w\sigma_u^2}\textbf{K}\textbf{K}^t$ and $\textbf{K}^t{\boldsymbol{\Sigma}^{\otimes}}^{-1} \textbf{K}= w\textbf{I}_n - w\delta \textbf{I}_n = (1-\delta)\textbf{W}$. And since in this balanced case $\widehat{\boldsymbol{\beta}_{\boldsymbol{\Lambda}}} = \widehat{\boldsymbol{\beta}_{ols}} = \widehat{\boldsymbol{\beta}_{w}}$, we have that $\left(\overline{\textbf{y}}-\textbf{X}\widehat{\boldsymbol{\beta}_{\boldsymbol{\Lambda}}}\right)^t\textbf{K}^t{\boldsymbol{\Sigma}^{\otimes}}^{-1}\textbf{K}\overline{\textbf{y}} = (1-\delta) Q_2$. Finally, see Eq. \ref{Eq:sigma_kron}, $\mid{\boldsymbol{\Sigma}^{\otimes}}\mid = (1-\delta)^{-n}\Rightarrow\mid{\boldsymbol{\Sigma}^{\otimes}}^{-1}\mid = (1-\delta)^{n}$ giving the marginal log likelihood for the simple balanced design in Eq. \ref{Eq:pdf_simple_mixed}.

\section{Appendix 2: More calculations on the quadratic form}\label{Ap:more_calc}
Furthermore we have that 
\begin{equation}
\begin{split}
\left(\overline{\textbf{y}}-\textbf{X}\widehat{\boldsymbol{\beta}_{\boldsymbol{\Lambda}}}\right)^t&\textbf{K}^t{\boldsymbol{\Sigma}^{\otimes}}^{-1}\textbf{K}\left(\overline{\textbf{y}}-\textbf{X}\widehat{\boldsymbol{\beta}_{\boldsymbol{\Lambda}}}\right)\\
= &\left(\overline{\textbf{y}}-\textbf{X}\widehat{\boldsymbol{\beta}_{w}}+\textbf{X}\widehat{\boldsymbol{\beta}_{w}}-\textbf{X}\widehat{\boldsymbol{\beta}_{\boldsymbol{\Lambda}}}\right)^t[\textbf{W}-\textbf{G}]\left(\overline{\textbf{y}}-\textbf{X}\widehat{\boldsymbol{\beta}_{w}}+\textbf{X}\widehat{\boldsymbol{\beta}_{w}}-\textbf{X}\widehat{\boldsymbol{\beta}_{\boldsymbol{\Lambda}}}\right)\\
= &\left(\overline{\textbf{y}}-\textbf{X}\widehat{\boldsymbol{\beta}_{w}}\right)^t[\textbf{W}-\textbf{G}]\left(\overline{\textbf{y}}-\textbf{X}\widehat{\boldsymbol{\beta}_{w}}\right)\\
& + 2\overline{\textbf{y}}^t\left(\textbf{I}_n - \underset{\textbf{H}_w^t}{\underbrace{\textbf{WX}\left(\textbf{X}^t\textbf{WX}\right)^{-1}\textbf{X}^t}}\right)[\textbf{W}-\textbf{G}]\textbf{X}\left(\widehat{\boldsymbol{\beta}_{w}}-\widehat{\boldsymbol{\beta}_{\boldsymbol{\Lambda}}}\right)\\
&+ \overline{\textbf{y}}^t\left(\textbf{H}_w^t-\textbf{H}_{\boldsymbol{\Lambda}}^t\right)[\textbf{W}-\textbf{G}]\textbf{X}\left(\widehat{\boldsymbol{\beta}_{w}}-\widehat{\boldsymbol{\beta}_{\boldsymbol{\Lambda}}}\right)\\
= &\left(\overline{\textbf{y}}-\textbf{X}\widehat{\boldsymbol{\beta}_{w}}\right)^t[\textbf{W}-\textbf{G}]\left(\overline{\textbf{y}}-\textbf{X}\widehat{\boldsymbol{\beta}_{w}}\right)\\& + 2\overline{\textbf{y}}^t\underset{= \textbf{0}_n\textbf{0}_n^t}{\underbrace{\left(\textbf{I}_n-\textbf{H}_{\boldsymbol{\Lambda}}^t\right)[\textbf{W}-\textbf{G}]\textbf{X}}}\left(\widehat{\boldsymbol{\beta}_{w}}-\widehat{\boldsymbol{\beta}_{\boldsymbol{\Lambda}}}\right)\\
& - 2\overline{\textbf{y}}^t\left(\textbf{H}_w^t-\textbf{H}_{\boldsymbol{\Lambda}}^t\right)[\textbf{W}-\textbf{G}]\textbf{X}\left(\widehat{\boldsymbol{\beta}_{w}}-\widehat{\boldsymbol{\beta}_{\boldsymbol{\Lambda}}}\right)\\
&+ \overline{\textbf{y}}^t\left(\textbf{H}_w^t-\textbf{H}_{\boldsymbol{\Lambda}}^t\right)[\textbf{W}-\textbf{G}]\textbf{X}\left(\widehat{\boldsymbol{\beta}_{w}}-\widehat{\boldsymbol{\beta}_{\boldsymbol{\Lambda}}}\right)\\
= &\left(\overline{\textbf{y}}-\textbf{X}\widehat{\boldsymbol{\beta}_{w}}\right)^t[\textbf{W}-\textbf{G}]\left(\overline{\textbf{y}}-\textbf{X}\widehat{\boldsymbol{\beta}_{w}}\right)\\
&- \left(\widehat{\boldsymbol{\beta}_{w}}-\widehat{\boldsymbol{\beta}_{\boldsymbol{\Lambda}}}\right)^t\textbf{X}^t[\textbf{W}-\textbf{G}]\textbf{X}\left(\widehat{\boldsymbol{\beta}_{w}}-\widehat{\boldsymbol{\beta}_{\boldsymbol{\Lambda}}}\right)\\
= &Q_2(1-\delta)^n -\text{trace}\left[\left(\widehat{\boldsymbol{\beta}_{w}}-\widehat{\boldsymbol{\beta}_{\boldsymbol{\Lambda}}}\right)\left(\widehat{\boldsymbol{\beta}_{w}}-\widehat{\boldsymbol{\beta}_{\boldsymbol{\Lambda}}}\right)^t\textbf{X}^t[\textbf{W}-\textbf{G}]\textbf{X}\right]
\end{split}
\end{equation}

\end{document}